# Disentangling stress and strain effects in ferroelectric HfO$_2$

Tingfeng Song,[1] Veniero Lenzi,[2] José P. B. Silva,[3,4] Luís Marques,[3,4] Ignasi Fina,[1,a)] Florencio Sánchez[1,a)]

[1]Institut de Ciència de Materials de Barcelona (ICMAB-CSIC), Campus UAB, Bellaterra 08193, Barcelona, Spain

[2]CICECO – Aveiro Institute of Materials, Department of Chemistry, University of Aveiro, 3810-193 Aveiro, Portugal

[3]Physics Center of Minho and Porto Universities (CF-UM-UP), University of Minho, Campus de Gualtar, 4710-057 Braga, Portugal

[4]Laboratory of Physics for Materials and Emergent Technologies, LapMET, University of Minho, 4710-057 Braga, Portugal

[a)]Authors to whom correspondence should be addressed: ifina@icmab.es and fsanchez@icmab.es

**Keywords:** Ferroelectric HfO$_2$; Oxides; Epitaxy; Lattice strain

**ABSTRACT:** Ferroelectric HfO$_2$ films are usually polycrystalline and contain a mixture of polar and nonpolar phases. This challenges the understanding and control of polar phase stabilization and ferroelectric properties. Several factors such as dopants, oxygen vacancies, or stress, among others, have been investigated and shown to have a crucial role on optimizing the ferroelectric response. Stress generated during deposition or annealing of thin films is a main factor determining the formed crystal phases and influences the lattice strain of the polar orthorhombic phase. It is difficult to discriminate between stress and strain effects on polycrystalline ferroelectric HfO$_2$ films, and the direct impact of orthorhombic lattice strain on ferroelectric polarization has yet to be determined experimentally. Here, we analyze the crystalline phases and lattice strain of several series of doped HfO$_2$ epitaxial films. We conclude that stress has a critical influence on metastable orthorhombic phase stabilization and ferroelectric polarization. On the contrary, the lattice deformation effects are much smaller than those caused by variations in the orthorhombic phase content. The experimental results are confirmed by density functional theory calculations on $HfO_2$ and $Hf_{0.5}Zr_{0.5}O_2$ ferroelectric phases.

**1. INTRODUCTION**

Ferroelectricity in $HfO_2$ occurs when this oxide crystallizes in the orthorhombic space group $Pca2_1$.[1] However, this is a high-energy polymorph and $HfO_2$ tends to stabilize in the centrosymmetric monoclinic phase (space group $P2_1/c$) of lower formation energy, thus the stabilization of polar orthorhombic phase is challenging.[2] Thin film crystallization involves high temperature deposition or annealing and kinetic factors are important, because the relative stability of polymorphs (monoclinic, orthorhombic and tetragonal $P4_2/nmc$ phases) differs with temperature, and a phase transformation requires surpassing an energy barrier.[3] In addition, the Gibbs energy of the competing polymorphs can be altered by several factors, which can favor the stabilization of the orthorhombic phase. In particular, partial substitution of Hf by another cation is critical. This is demonstrated for a broad number of doping atoms (Zr, Si, La, Y, etc.), being the optimal dopant concentration specific of each case.[4–7] Moreover, anion doping (as N, C or H) or oxygen vacancies can also reduce the Gibbs energy of the orthorhombic phase.[5,8] Nevertheless, doping is not enough to make the orthorhombic phase the lowest energy polymorph, and additional contributions are needed. Among them, the relevance of surface and interface energy is evidenced by the usual reduction or vanishment of ferroelectric polarization in films thicker than a few tens of nanometers.[9–12]

Besides, stress effects are relevant. Stress in a polymorph increases its energy either directly by inducing elastic strain in sufficiently thin films or indirectly by creating dislocations or other defects. The increased energy can be very different among the competing polymorphs, and this can help a metastable phase to stabilize. In the case of polycrystalline doped $HfO_2$ films, tensile stress during cooling after crystallization of films on substrates with small thermal expansion coefficient results in high polarization.[13] In addition, stress during crystallization induced by a top electrode can have also an important effect.[6,14]

In the case of epitaxial films deposited at high temperature, crystallization occurs almost immediately as atoms arrive to the surface of the substrate,[15] and the lattice mismatch between the crystallized $HfO_2$ film and the substrate causes stress. It is demonstrated that this stress has a huge effect on the stabilization of the $HfO_2$ phases, and the amount of orthorhombic phase can be maximized by using proper substrate [16,17] and deposition parameters.[18]

As discussed above, cation and/or anion doping and stress are important factors on the stabilization of the orthorhombic crystal phase against the other polymorphs. Furthermore, doping can cause chemical strain in the orthorhombic lattice, and this chemical strain could influence the ferroelectric properties, similar to how elastic strain does. Additional strain in the orthorhombic lattice can occur due to point-like defects,[19] and also by elastic coupling between coexisting phases in the film[20,21] or by bending if the film is deposited on (or transferred to) a flexible substrate.[22] Lattice strain can influence the magnitude and orientation of the atomic dipoles and thus can have a direct influence on the polarization of a ferroelectric.[23,24] This is usually observed in $BaTiO_3$ and other ferroelectric perovskites.[25–27] In the case of ferroelectric $HfO_2$-based compounds, theoretical studies of the effect of dopants, oxygen vacancies, surface energy, and stress on the relative energy of the polymorphs are reported.[7,28–32] For instance, Materlik et al.[28] concluded that the stabilization of the $Pca2_1$ phase could not be achieved only by strain-induced stress, because a large compressive biaxial stress would be needed, and therefore surface energy contributions must play an important role. This seems to be the case for (111)-oriented films as well.[33] Later studies showed that the $P2_1/c$ phase might be rendered less favorable than the $Pca2_1$ phase by mechanical constraints, such as those induced in electrode capping or substrate clamping.[34] Further, it was recently suggested by high-throughput density functional theory (DFT) calculations that the $Pca2_1$ phase can be stabilized under wide range of epitaxial conditions achievable on substrates such as YSZ or STO, in which symmetry constraint-induced strains could suppress the formation of $P2_1/c$ phase.[35]

The direct effect of lattice strain on the polarization of the orthorhombic phase is less investigated. It has been predicted slight increase (reduction) of polarization with biaxial compressive (tensile) strain.[36] In this regard, Wei et al.[37] investigated lattice strain effects on polarization and switching energy barrier considering two possible polarization switching pathways in orthorhombic phase $HfO_2$ (depending on whether the three-coordinated oxygen atoms either cross or not the hafnium planes). Their calculations indicate that effects were highly different if strain was uniaxial or biaxial and, depending on the pathway, an increase in polarization could occur under tensile or under compressive strain. In the case of epitaxial films, Liu et al.[34] calculated the effect of epitaxial strain for (111) oriented $HfO_2$ films and found a monotonic increase (reduction) of polarization from compressive (tensile) stress.

Lattice parameters of polycrystalline doped $HfO_2$ films are usually determined by X-ray diffraction (XRD) and particularly from the analysis of the orthorhombic (111) Bragg peak, which position is very close to that of the tetragonal (101) and cubic (111) peaks. Peak shape and position are generally evaluated to quantify the ratio of orthorhombic and tetragonal phases,[38] without considering elastic strain, homogeneous or inhomogeneous, in the orthorhombic phase. The mentioned Bragg peaks are very close, and diffraction peaks are broad due to the ultrathin character of the films, which challenges disentangling phase's ratio and strain of the orthorhombic phase in polycrystalline films. In contrast, in doped $HfO_2$ epitaxial films on $La_{0.67}Sr_{0.33}MnO_3$ (LSMO) electrodes, transmission electron microscopy characterization and the correlation between ferroelectric polarization and the intensity of the mentioned diffraction peak points to absence of tetragonal phase.[16,18,39] Thus, epitaxial films of doped $HfO_2$ on LSMO are appropriate systems to determine if lattice strain of the orthorhombic phase has a direct influence on the ferroelectric polarization. In the case of conventional ferroelectric perovskites as $BaTiO_3$, this is usually discerned by tuning the strain through epitaxial growth on various substrates with different lattice parameters.[23,24] However, this method is not suitable for ferroelectric doped $HfO_2$ films on LSMO due to the change in phase ratio with the substrate[16,17] and because the domain matching epitaxy mechanism of $HfO_2$ on LSMO results in low elastic strain.[40] Aiming at further discerning between stress and strain contributions, we compare here the formed phases and the ferroelectric polarization of more than 50 epitaxial films of doped $HfO_2$. Epitaxial $Hf_{0.5}Zr_{0.5}O_2$ (HZO), La-doped (1 at.%) HZO, and La-doped (2 at.%) $HfO_2$ (HLO) films were grown by pulsed laser deposition (PLD) on LSMO electrodes (see Experimental Section and Table S1 in the supplementary material for a description of the series of films). The study includes series of HZO and HLO films grown on oxide substrates with various lattice parameters and on Si(001), four series of HZO films grown under different conditions of fixed Ar pressure (0, 0.05, 0.1 and 0.2 mbar) with varied $O_2$ pressure, and series of HZO films grown under different conditions of ambient pressure (Ar and $O_2$) and temperature. The investigation of a broad range of epitaxial thin films with different doping and deposition parameters allows disclosing the governing mechanisms for achieving robust ferroelectric polarization in the films. Therefore, in the present work, correlations between polarization and both orthorhombic phase content and orthorhombic lattice strain along [111] are established. The content of the orthorhombic phase determines the polarization, while there are not evidences of a strong direct effect of the strain on the polarization.

## 2. RESULTS AND DISCUSSION

Fig. 1a shows a sketch of the epitaxial heterostructures and a selection of XRD 2θ-χ maps. The phases of HZO films grown on substrates with increasing lattice parameters range from purely monoclinic to purely orthorhombic, and there is a mixture of both phases in films on substrates with intermediate lattice parameter. Very recently, Song et al.[17] reported the effect of the epitaxial stress on La-doped $HfO_2$ films (Fig. 1b) and showed that increasing the lattice parameter of the substrate the films evolved from purely monoclinic to purely orthorhombic. However, the metastable orthorhombic phase is formed on substrates with smaller lattice parameter in HLO films than in HZO films. The XRD 2θ-χ maps in Fig. 1a and 1b illustrate this difference. A HZO film on $NdGaO_3$ (pseudocubic lattice parameter $a_s$ = 3.86 Å) is mostly monoclinic, with minor content of orthorhombic phase. In contrast, in the HLO film grown on the same substrate, the orthorhombic phase is the main phase and the monoclinic is secondary. The stress effects also differ in films grown on $SrTiO_3$ ($a_s$ = 3.905 Å). HZO is mostly orthorhombic but there is high amount of monoclinic phase, while the monoclinic phase is not detected in the HLO film. Finally, HZO and HLO films on substrates with larger lattice parameter as $GdScO_3$ ($a_s$ = 3.973 Å) are single orthorhombic.

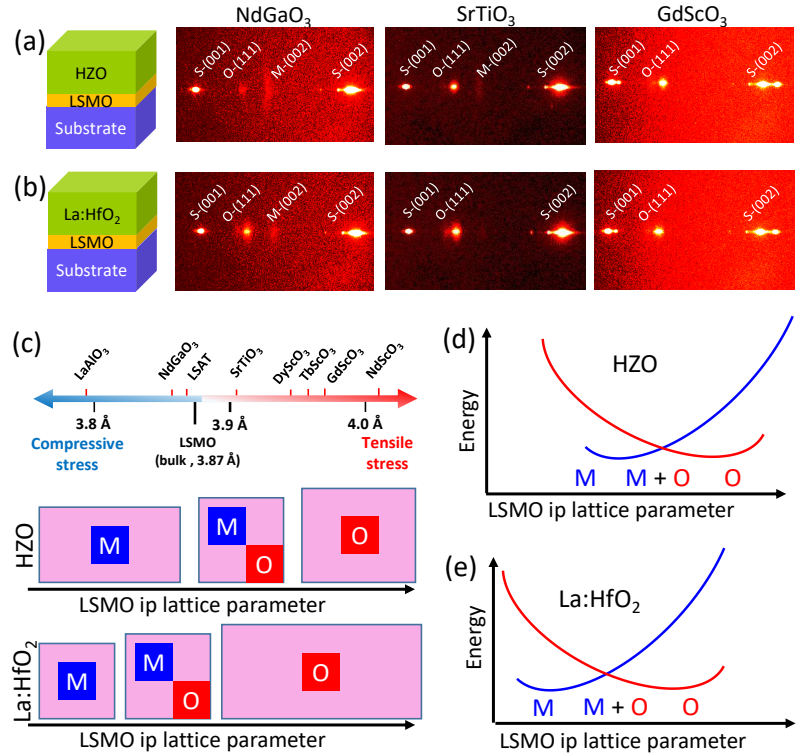

**FIG. 1**. XRD χ (horizontal axis) - 2θ (vertical axis) maps of HZO (a) and HLO (b) films on NdGaO$_3$, SrTiO$_3$ and GdScO$_3$ substrates. Labels S-(001), O-(111) and M-(002) mark the 2θ position of the (001), (111) and (002) reflections of the substrate, orthorhombic phase and monoclinic phase, respectively. (c) Sketch indicating the stress on the LSMO electrode, compressive or tensile, caused by several oxide substrates, and the impact on the phases formed in HZO and LHO films. (d) and (e) Sketch of the impact of the in-plane (ip) lattice parameter of the LSMO electrode on the energy of monoclinic and orthorhombic HZO and HLO films, respectively.

Fig. 1c sketches the critical role of the epitaxial stress on the crystal phase, and the important differences between HZO and HLO. The orthorhombic phase is stabilized on a broader set of substrates in HLO than in HZO, which means that the energy is lower when the orthorhombic phase forms on these substrates instead of the monoclinic phase. The low interface energy is key factor in epitaxial stabilization of metastable phases,[41] and thus the orthorhombic phase is perhaps directly stabilized when the film crystallizes epitaxially at high temperature. In this case, epitaxial films of doped $HfO_2$ would not undergo the phase transformations that occur in polycrystalline films.[3] Assuming this hypothesis, the formation of monoclinic and orthorhombic phases in HZO and HLO films on the different substrates could be understood considering only thermodynamics, as sketched in Fig. 1d and 1e for HZO and HLO, respectively. The thickness of the HZO and HLO films is similar (~9.5 and 9 nm, respectively) and the films were grown using same deposition parameters.[16,17] Thus, the contributions to the total energy that could favor the stabilization of the orthorhombic phase are the bulk energy and the $HfO_2$ film/LSMO interface energy. The bulk energy of the orthorhombic phase is greater than the monoclinic phase, but the difference can depend on the dopant atom.[7,42,43] A smaller difference in the bulk energy between both phases would favor the stabilization of the orthorhombic phase on substrates with smaller lattice parameter. However, it would also make the orthorhombic phase more stable increasing thickness in HLO films than in HZO films, and this does not occur. Fig. S1 in the supplementary material shows the dependence on thickness of the remanent polarization ($P_r$) of HZO[18] and HLO[12] films on LSMO/$SrTiO_3$(001) substrates. It is seen that for films thinner than 10 nm, when surface and interface contributions to the total energy are dominant, $P_r$ is higher in HLO than in HZO films. However, for thicker films, when bulk energy is the main contribution to the total energy, $P_r$ is similar for HZO and HLO films of same thickness. Thus, we discard differences in bulk energy as a main factor to favor the observed stabilization of the orthorhombic phase. The second contribution that can explain the greater amount of orthorhombic phase in HLO films on substrates with smaller lattice parameter is the energy of the $HfO_2$/LSMO interface. The present results suggest that the energy of the monoclinic phase/LSMO interface is higher than that of the orthorhombic phase/LSMO interface on these substrates, thus favoring the epitaxial stabilization of the orthorhombic phase.

The ferroelectric polarization, the intensity of the orthorhombic (111) Bragg peak and the $d_{(111)}$ interplanar spacing of the epitaxial HZO[16] and HLO[17] films depend on the substrate and deposition parameters.[18] The normalized intensity of the orthorhombic (o) (111) peak and the $d_{(111)}$ spacing of films grown under same deposition parameters on different substrates are plotted as a function of the remanent polarization in Fig. 2. The o-(111) peak intensity and the $d_{(111)}$ spacing permit quantifying the content and strain of the orthorhombic phase, respectively. There is a clear correlation between increasing the peak intensity and the ferroelectric polarization, in agreement with previous studies done with polycrystalline doped $HfO_2$ films.[44,45] The $d_{(111)}$ spacing of all the HLO films is very similar, in the 2.978 - 2.987 Å range,[17] while, in the case of the HZO films, the $d_{(111)}$ spacing is smaller than that of HLO films and ranges from 2.948 to 2.975 Å, having the films on Nd, Gd and Tb scandate substrates the smallest values.[16] We note that HZO $d_{(111)}$ spacing reported in ref. [16] was determined using a XRD two-dimensional detector and the accuracy was limited, and most of the found variations are below the accuracy limit. XRD θ-2θ measurements using point detector show more similar $d_{(111)}$ spacing in the HZO films, being about 2.973 Å on MgO and about 2.96 Å in others samples with detectable peak. Thus, it can be appreciated that whereas the HZO and HLO films have very similar $d_{(111)}$ spacing of about 2.98 Å, the polarization changes significantly, from almost null $P_r$ to near 30 μC cm$^{-2}$. On the other hand, some films with similar $P_r$ of 20-25 μC cm$^{-2}$ have $d_{(111)}$ spacing ranging from about 2.95 to about 2.99 Å. In summary, from Fig. 2 it can be concluded that the substrate selection defines the orthorhombic phase content, which is the major parameter controlling the ferroelectric polarization in the films. Moreover, the huge differences in polarization among the films are not correlated with the $d_{(111)}$ spacing value.

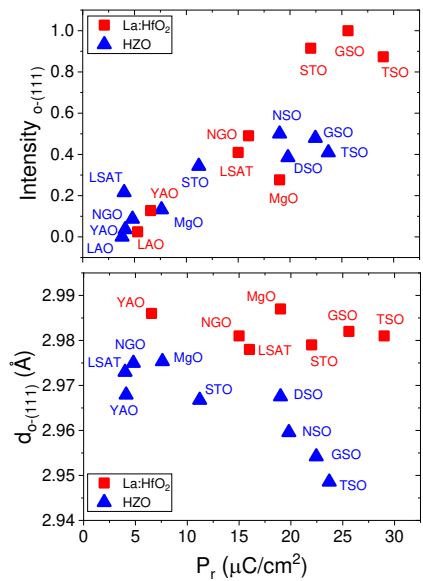

**FIG. 2**. Top panel: Normalized intensity of the orthorhombic (111) diffraction peak of HZO (blue triangles) and HLO (red squares) films deposited on various substrates: $YAlO_3$

(YAO), LaAlO$_3$ (LAO), NdGaO$_3$ (NGO), La$_{0.18}$Sr$_{0.82}$Al$_{0.59}$Ta$_{0.41}$O$_3$ (LSAT), SrTiO$_3$ (STO), DyScO$_3$ (DSO), TbScO$_3$ (TSO), GdScO$_3$ (GSO), NdScO$_3$ (NSO) and MgO substrates. Bottom panel: corresponding d$_{(111)}$ spacing plotted as a function of the remanent polarization of the films. Data plotted here were reported in ref. [16,39].

Besides stress engineering by substrate selection, modification of growth parameters also permits controlling the amount of orthorhombic phase. In particular, the oxygen partial pressure[18] and the combination of O$_2$ and Ar gases[19] in PLD have a critical impact. The combination of O$_2$ and Ar allows the low oxidation conditions that favor the stabilization of the orthorhombic phase, while avoiding the very high energy of the PLD plasma that can degrade film crystallinity. Fig. 3 shows the dependence of P$_r$ with the content of orthorhombic phase for films deposited on SrTiO$_3$(001) at substrate temperature T$_S$ = 800 °C and with different combinations of oxygen and argon: P$_{Ar}$ = 0 mbar, P$_{Ar}$ = 0.005 mbar, P$_{Ar}$ = 0.1 mbar and P$_{Ar}$ = 0.2 mbar. For all series, P$_r$ is higher with increasing orthorhombic phase content, with linear-like dependence in each series (Fig. 3, inset). The exception observed in the P$_{Ar}$ = 0.2 mbar samples is because the larger leakage contribution of one of the two films (empty down triangle), which does not permit obtaining reliable P$_r$ values. We note that XRD peak intensity increases with the thickness of the epitaxial HfO$_2$ films while polarization usually diminished, and thus thickness difference among samples may cause some scattering in the data. To diminish fluctuations due to thickness variation, the ferroelectric polarization and XRD intensity data of the samples, shown in Fig. 3, have been extrapolated in Fig. S2 (supplementary material) to values corresponding to a thickness of 10 nm according thickness dependences reported in ref. [18]. The normalization does not alter the linear dependence between polarization and orthorhombic peak intensity, and indeed, the data of the different series are closer.

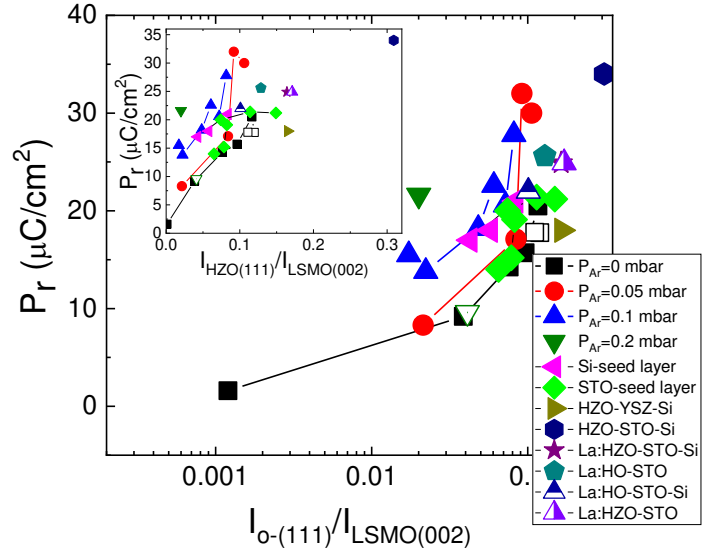

**FIG. 3**. Remanent polarization as a function of the intensity (logarithmic scale) of the orthorhombic (111) diffraction peak, normalized to the LSMO(002) peak intensity. The graph includes four series of films deposited on $SrTiO_3$(001) with various $P_{O2}$ and $P_{Ar}$; two other series were deposited at various temperatures on $SrTiO_3$(001) and Si(001) previously coated with seed layers. Data of films on $SrTiO_3$ and YSZ buffered Si are included, as well as La-doped $HfO_2$ and La-doped HZO on $SrTiO_3$(001) and Si(001). Empty symbols: samples with underestimated polarization due to leakage. See legend in the graph and Experimental Section for sample identification and further details. Inset: Same plot with the intensity of the diffraction peak in linear scale.

The deposition temperature, $T_S$, is another deposition parameter that impacts orthorhombic phase content, strain, and ferroelectric polarization. The orthorhombic phase can be stabilized in a broad range of $T_S$ through growth on a seed layer.[46] Fig. 3 includes data of films grown epitaxially in the $T_S$ = 550 - 800 °C range under $P_{O2}$ = 0.1 mbar and $P_{Ar}$ = 0 mbar. The polarization of these films exhibits the same linear dependence with the orthorhombic phase content. Polarization and orthorhombic phase content of HLO and La-doped (1 at.%) HZO films deposited on $SrTiO_3$(001) at 800 °C with $P_{O2}$ = 0.1 mbar and $P_{Ar}$ = 0 mbar are also shown. It is evidenced that different strategies (here selection of the dopant atom, deposition temperature, or deposition ambient pressure) permit tailoring the fraction of orthorhombic phase, and this fraction determines the ferroelectric polarization. Fig. 2 showed that the selection of a particular oxide substrate also has a similar effect. In Fig. 3 we have added data of HLO and La-doped (1 at.%) $HfO_2$ films on Si(001) substrates buffered with $SrTiO_3$ [12,47] or YSZ[48] layers. The graph of polarization as a function of the intensity of the orthorhombic (111) peak (Fig. 3) and the equivalent graph for films on various oxide substrates (Fig. 2a) are combined in Fig. S3 in the supplementary material. Since normalization of XRD intensity to the LSMO(002) peak was not done in Fig. 2a due to overlapping with substrate reflection, in Fig. S3 in the supplementary material the intensity of the orthorhombic (111) peak is normalized to the La:$HfO_2$ film on $SrTiO_3$(001) present in both Fig. 2a and Fig. 3. Fig. S3 in the supplementary material, showing data of more than 50 samples and despite the strong and opposite dependence that polarization and XRD intensity have on $HfO_2$ thickness, confirms that polarization is mainly determined by the content of orthorhombic phase.

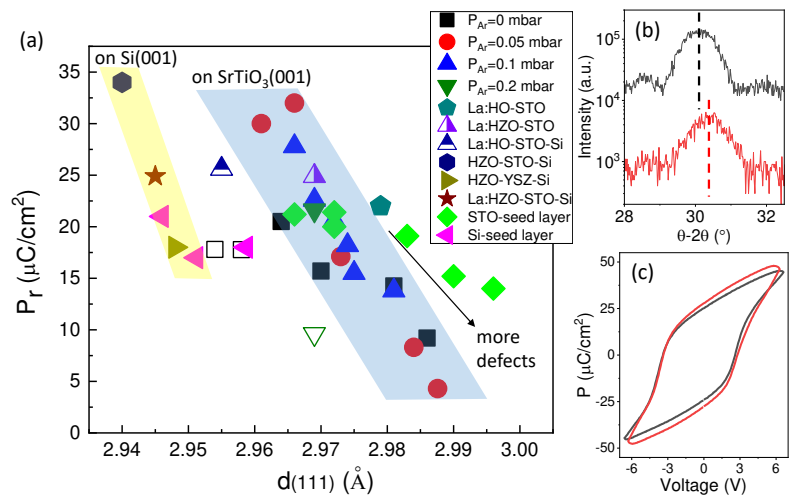

**FIG. 4**. (a) Remanent polarization as a function of $d_{(111)}$ spacing for same samples shown in Fig. 3 (same symbols are used here). The graph includes four series of films deposited on $SrTiO_3$(001) with various $P_{O2}$ and $P_{Ar}$; two other series were deposited at various temperatures on $SrTiO_3$(001) and Si(001) previously coated with seed layers. Data of films on $SrTiO_3$ and YSZ buffered Si are included, as well as La-doped $HfO_2$ and La-doped HZO on $SrTiO_3$(001) and Si(001). Empty symbols: samples with underestimated polarization due to leakage. See legend in the graph and Experimental Section for sample

identification and further details. Blue and yellow colored regions mark the $P_r$ - $d_{(111)}$ trend for films on SrTiO$_3$(001) and Si(001), respectively. The thickness of the films is indicated in Table S2 in the supplementary material. (b) XRD θ-2θ scans and (c) polarization loops of La:HZO on SrTiO$_3$(001) (black line) and Si(001) (red line).

Fig. 3 confirms the direct impact of the orthorhombic phase content on the polarization. Next, we evaluate (Fig. 4) if lattice strain has a direct effect. The $d_{(111)}$ spacing is in the 2.96 - 2.99 Å range in most of the HZO and HLO films on SrTiO$_3$(001), and films with shorter $d_{(111)}$ tend to have higher $P_r$ (marked with cyan color). Some films (open symbols) on SrTiO$_3$(001) are leaky and their $P_r$ was underestimated, which shifts their $P_r$ - $d_{(111)}$ data out of the marked cyan region. The $P_r$ - $d_{(111)}$ correlation in the marked cyan region can suggest a huge direct effect of strain on polarization, since $P_r$ ranges from almost null value to about 30 μC cm$^{-2}$ when $d_{(111)}$ reduces moderately from 2.99 to 2.96 Å. However, this would be contradictory with the comparison of films on different oxide substrates (Fig. 2), that discarded a highly important direct effect of strain. We note that the larger $d_{(111)}$ in Fig. 4a correspond to films grown under low total (including both O$_2$ and Ar) pressure or deposited at low $T_S$, which likely present high density of point defects that expand the crystal cell and degrade the ferroelectric behavior. This causes a correlation (Fig. S4 in the supplementary material) between polarization and strain, but without implying a significantly high direct effect of $d_{(111)}$ strain on polarization. Indeed, the comparison of films on SrTiO$_3$(001) (cyan region) and Si(001) (yellow region) further demonstrates that $d_{(111)}$ strain has not an important effect on polarization. The thermal expansion coefficient (TEC) of Si is much smaller than that of SrTiO$_3$, producing strong tensile stress to the doped HfO$_2$ films during cooling after growth at high temperature. The TEC mismatch causes an important reduction of the out-of-plane $d_{(111)}$ lattice spacing, and $d_{(111)}$ values of films on Si(001) are in the 2.94-2.95 Å range. This strain, produced after crystallization, shifts the $P_r$ - $d_{(111)}$ data of films on Si(001) to the left in Fig. 4a, and confirms that a change in $d_{(111)}$ does not affects significantly the polarization. The comparison of XRD patterns (Fig. 4b) and polarization loops (Fig. 4c) of La:HZO films on STO (black lines) and Si(001) (red lines) illustrate the limited effect of strain. It has to be noted that the ferroelastic transitions that occur in orthorhombic HfO$_2$ induced either by stress[49] or electric field[50] are not observed in our films due to the (111) orientation. Recently, Zhong et al.[22] fabricated epitaxial HZO(111) membranes and reported negligible influence of strain provided by bending on the ferroelectric polarization, in agreement with our results. Although strain effects could be reduced by the (111) texture of the films, the results demonstrate that polarization in ferroelectric $HfO_2$ is directly influenced by the content of orthorhombic phase, which masks less relevant strain effects.

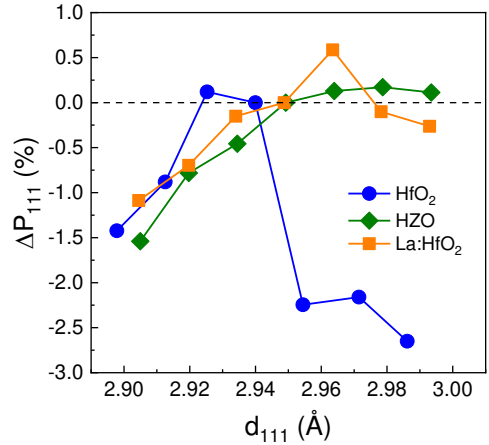

**FIG. 5**. Density functional theory calculations of the spontaneous polarization $P_{111}$ in the (111) direction as a function of $d_{111}$ for the orthorhombic (111)-oriented phase of $HfO_2$, (blue circles) HZO (green diamonds) and La:$HfO_2$ (orange squares). Results are reported as relative difference with respect to the $P_{111}$ value of the fully relaxed and strain-free unit cells.

The experimental observations have been confirmed by theoretical calculations. We investigated the effect of the variations of $d_{111}$ on the ferroelectric polarization in $HfO_2$, La:$HfO_2$ and HZO. The calculated polarization as a function of $d_{(111)}$ is shown in Fig. 5. We considered an interval of strains between -1.5 % and 1.5 %, which reproduces the experimentally observed $d_{111}$ values. For all systems, our calculations show that changes in $d_{111}$ have a negligible impact on the ferroelectric polarization, with a variation not greater than 3% for all the strain values considered. Thus, the calculations demonstrate that the out-of-plane strain effects do not alter significantly the ferroelectric polarization at the experimentally observed values.

**3. CONCLUSIONS**

In summary, using epitaxial films as a model system, we have performed a comprehensive study of the stabilization of the orthorhombic phase of doped $HfO_2$, its lattice strain, and its ferroelectric polarization. The stabilization of the orthorhombic phase of HLO on a broader range of substrates than HZO is due to differences in interface energy and not to changes in the relative bulk energy of polymorphs between HLO and HZO. The selection of the substrate, the dopant atom and the deposition parameters allow tuning the fraction of orthorhombic phase, and this fraction determines the polarization. Deposition parameters also influence strain, and the correlation with orthorhombic phase content challenges disentangling the impact of both contributions on polarization. This difficulty is overcome comparing films on different substrates. The huge variation in polarization observed in equivalent (same composition and deposition conditions) films on different oxide substrates and presenting same strain is correlated with the orthorhombic phase content. In the case of films on Si(001), the high compressive stress due to the low thermal expansion coefficient of silicon causes important strain in the film. The DFT calculations confirm that the out-of-plane strain of the films has a very small influence on the polarization. In conclusion, with this work we clarify the role of stress and strain in $HfO_2$-based thin films: 1) stress has a critical impact on the crystallizing phase between competing polymorphs and therefore on the polarization, and 2) lattice strain, contrary to conventional ferroelectrics, has very little effect on polarization.

**4. EXPERIMENTAL SECTION**

The ferroelectric $HfO_2$ film and the LSMO electrode were deposited in a single process by PLD using a KrF excimer laser. Two series of HZO and HLO films deposited on various oxide single crystalline substrates at 800 °C substrate temperature and 0.1 mbar oxygen partial pressure. Additional experimental details and properties of the films are reported in ref. [16,17]. Other HZO and HLO films, deposited at $T_S$ = 800 °C and $P_{O2}$ = 0.1 mbar, were integrated epitaxially with Si(001), using yttria-stabilized zirconia (YSZ)[48] or $SrTiO_3$ (STO)[12,51] buffer layers. La-doped HZO films were also deposited at $T_S$ = 800 °C and $P_{O2}$ = 0.1 mbar on $SrTiO_3$(001) and STO-buffered Si(001) substrates.[52] Four series of HZO films were deposited at $T_S$ = 800 °C on $SrTiO_3$(001) substrates under mixed Ar and $O_2$ pressure (fixed $P_{Ar}$ = 0 mbar and varied $P_{O2}$ from 0.01 to 0.2 mbar, fixed $P_{Ar}$ = 0.05 mbar and varied $P_{O2}$ from 0.005 to 0.1 mbar, fixed $P_{Ar}$ = 0.1 mbar and varied $P_{O2}$ from 0.002 to 0.1 mbar, and fixed $P_{Ar}$ = 0.2 mbar and $P_{O2}$ of 0.01 and 0.05 mbar). The range of total pressure in the four series is 0.01 - 0.2 mbar, 0.055 - 0.15 mbar, 0.102 - 0.2 mbar and 0.21 - 0.25 mbar, respectively.[19] Finally, a series of HZO films was grown epitaxially on SrTiO$_3$(001) P$_{O2}$ = 0.1 mbar at various substrate temperature in the 550 - 800 °C range through the use of a seed layer.[46] Table S1 in the supplementary material summarizes the samples included in this study.

Structural characterization was performed by XRD using Cu Kα radiation. Circular platinum top electrodes (thickness 20 nm and diameter 20 μm) were deposited by dc magnetron sputtering through stencil masks for electrical characterization. Ferroelectric polarization loops were measured in pristine state at frequency of 1 kHz in top-bottom configuration (grounding the bottom electrode and biasing the top one) at room temperature using an AixACCT TFAnalyser2000 platform. All epitaxial films of this study have not wake-up effect when are cycled in saturation condition.[53]

Density functional theory calculations were performed using the Vienna Ab initio Simulation Package (VASP).[54–56] The PBEsol[57] functional and an energy cutoff of 600 eV was used in all calculations. For pure HfO$_2$, a (111)-oriented orthorhombic phase unit cell containing 36 atoms was considered along with a 4×4×3 k-point grid. To model HZO we built a special quasirandom structure (SQS)[58] for a 144-atom (111)-oriented supercell and employed a 2×2×3 k-point grid. The La doped HfO$_2$ was obtained by introducing a single La$_{Hf}$ substitution in a 144-atom (111)-oriented supercell, resulting in a doping concentration of 2 Hf at. %. To study the effect of the change of d$_{(111)}$ interplanar spacing on polarization, the c lattice parameters of HfO$_2$, HZO and La:HfO$_2$ cells were strained and kept fixed while the in-plane lattice parameters were relaxed along with the atomic positions. Forces and stresses were relaxed up to a threshold of 2 meV Å$^{-1}$ and 0.1 GPa, respectively. The modern theory of polarization was used to calculate ferroelectric polarization on the relaxed structures.[59]

**SUPPLEMENTARY MATERIAL**

See the supplementary material for the dependence on thickness of polarization of Hf$_{0.5}$Zr$_{0.5}$O$_2$ and La (2%):HfO$_2$, the dependence of normalized (t=10nm) remanent polarization on normalized (t=10nm) o-(111) peak intensity, the remanent polarization as a function of the o-(111) intensity, the intensity of o-(111) peak as a function of the o-(111) out-of-plane lattice parameter, a table that summarizes the samples included in this study, and table indicating the thickness of the films shown in Fig. 4,.

**ACKNOWLEDGEMENTS**

Financial support from the Spanish Ministry of Science and Innovation (MCIN/AEI/ 10.13039/501100011033), through the Severo Ochoa FUNFUTURE (CEX2019-000917-S), PID2020-112548RB-I00 and PID2019-107727RB-I00 projects, from Generalitat de Catalunya (2021 SGR 00804) and from CSIC through the i-LINK (LINKA20338) program is acknowledged. We also acknowledge project TED2021-130453B-C21, funded by MCIN/AEI/10.13039/501100011033 and European Union NextGeneration EU/PRTR. TS is financially supported by China Scholarship Council (CSC) with No. 201807000104. This work was supported by: (i) the Portuguese Foundation for Science and Technology (FCT) in the framework of the Strategic Funding Contract UIDB/04650/2020; (ii) the exploratory research project 2022.01740.PDTC (DOI:10.54499/2022.01740.PTDC) and (iii) the project M-ERA-NET3/0003/2021 - NanOx4EStor grant agreement No 958174 (DOI:10.54499/M-ERA-NET3/0003/2021). J. P. B. S. also thanks FCT for the contract under the Institutional Call to Scientific Employment Stimulus – 2021 Call (CEECINST/00018/2021). This work was developed within the scope of the project CICECO-Aveiro Institute of Materials, UIDB/50011/2020, UIDP/50011/2020 & LA/P/0006/2020, financed by national funds through the FCT/MCTES (PIDDAC).